\documentclass[twocolumn]{emulateapj}

\usepackage{natbib} 
\usepackage{epsfig} 
\usepackage{lscape} 
\usepackage{graphicx} 
\slugcomment{To Appear in ApJ}

\shorttitle{{\it Spitzer} 24$\mu$m Time-Series Observations of GU Boo}

\shortauthors{von Braun et al.}

\begin{document}

\title{{\it Spitzer} 24\micron\ Time-Series Observations of
the Eclipsing M-dwarf Binary GU Bo\"otis }

\author{Kaspar von Braun\altaffilmark{1}}
\author{Gerard T. van Belle\altaffilmark{1,4}}
\author{David R. Ciardi\altaffilmark{1}}
\author{Mercedes L\'{o}pez-Morales\altaffilmark{2}}
\author{D.W. Hoard\altaffilmark{3}}
\author{Stefanie Wachter\altaffilmark{3}}

\altaffiltext{1}{Michelson Science Center, California Institute of
Technology, MC 100-22, Pasadena, CA 91125;
kaspar, gerard, ciardi@ipac.caltech.edu}

\altaffiltext{2}{Carnegie Fellow, Department of Terrestrial Magnetism, 
Carnegie Institution of Washington, 5241 Broad Branch Rd. NW,
Washington, DC 20015; mercedes@dtm.ciw.edu}

\altaffiltext{3}{{\it Spitzer} Science Center, California Institute of
Technology, MC 220-6, Pasadena, CA 91125;
hoard, wachter@ipac.caltech.edu}

\altaffiltext{4}{European Southern Observatory, Karl-Schwarzschild-Str. 2,
85748 Garching, Germany; gerard.van.belle@eso.org}


\begin{abstract}

We present a set of {\it Spitzer} 24$\mu$m MIPS time series observations of
the M-dwarf eclipsing binary star GU Bo\"otis. Our data cover three secondary
eclipses of the system: two consecutive events and an additional eclipse six
weeks later. The study's main purpose is the long wavelength (and thus limb
darkening-independent) characterization of GU Boo's light curve, allowing for
independent verification of the results of previous optical studies. Our
results confirm previously obtained system parameters. We further compare GU
Boo's measured 24$\mu$m flux density to the value predicted by spectral
fitting and find no evidence for circumstellar dust. In addition to GU Boo, we
characterize (and show examples of) light curves of other objects in the field
of view. Analysis of these light curves serves to characterize the photometric
stability and repeatability of {\it Spitzer's} MIPS 24\micron\ array over
short (days) and long (weeks) timescales at flux densities between
approximately 300--2,000$\mu$Jy. We find that the light curve root mean square
about the median level falls into the 1--4\% range for flux densities higher
than 1mJy. Finally, we comment on the fluctuations of the 24\micron\ background
on short and long timescales.

\end{abstract}


\keywords{techniques: photometric, infrared: stars, stars: fundamental 
parameters, stars: binaries: eclipsing, stars: individual: GU Boo,
circumstellar matter, dust}


\section{Introduction}\label{introduction}

GU Bo\"otis is a nearby, low-mass eclipsing binary system, consisting of two
nearly equal mass M-dwarfs \citep{lmr05}. It is one of currently very few
($\sim$5) known nearby ($<$ 200 pc) double-lined, detached eclipsing binary
(DEB) systems composed of two low-mass stars \citep{lm07}.  Eclipsing binaries
can be used as tools to constrain fundamental stellar properties such as mass,
radius, and effective temperature. Given the fact that over 70\% of the stars
in the Milky Way are low-mass objects with $M< 1 M_\odot$ \citep{hik97},
coupled with the considerable uncertainty over the mass-radius relation for
low-mass stars, objects such as GU Boo are of particular interest in exploring
the low-mass end of the Hertzsprung-Russell diagram.

While simultaneous analysis of DEB light curves and radial velocity (RV)
curves provides insight into the component masses and physical sizes, an
estimate of their intrinsic luminosities can only be made with the knowledge
of the amount and properties of dust along the line of sight. As such, it is
important to understand whether low-mass DEB systems used to constrain stellar
models contain dust which, in turn, may lead to an underestimate of their
surface temperatures and thus luminosities. This problem has been documented
in \citet{dfm99,mlg01,tr02,ribas03}. In particular, \citet{ribas03} states
that the most likely explanation for the temperature discrepancy between
observations and models for the low-mass DEB CU Cancri is the presence of
either circumstellar or circumbinary dust.  The detection of dust in any
system such as GU Boo would therefore additionally shed insight into formation
and evolution of the low-mass DEBs.

The characterization of the effects of limb darkening and star spots
introduces additional free parameters and thus statistical uncertainty in the
calculation of the stellar radii and masses.  Using the {\it Spitzer Space
Telescope}, we obtained 24$\mu$m time series observations of three separate
instances of GU Boo's secondary eclipse (see \S \ref{observations}) to create
a light curve far enough in the infrared to not be contaminated by the effects
of limb darkening and star spots. We purposely timed the observations such
that each secondary eclipse event is preceded by a sufficient length of time
to establish GU Boo's out-of-eclipse flux density in order to detect any
infrared excess possibly caused by thermal dust emission.

A further goal of our study is to characterize the photometric stability of
the Multiband Imaging Photometer (MIPS) on {\it Spitzer} at 24$\mu$m over
short and long time scales, similar to what was done for bright objects in \S
5 of \citet{rye04}. Time-series observing is atypical (albeit increasingly
common) for {\it Spitzer}, which is the reason why there are very few
published photometric light curves based on {\it Spitzer} observations. The
recent spectacular observations of primary and secondary eclipses of
transiting planets are notable exceptions \citep[see for
instance][]{cam05,dsr05,cac07,dhl07,gdb07,kca07}. Of these, the \citet{dsr05}
study was performed at 24$\mu$m. We therefore observed two consecutive
secondary eclipses of GU Boo ($\sim$ 12 hours apart), and then a third event
about six weeks later (see Table \ref{table1_AORs}).

We describe our observations and data reduction methods in \S
\ref{obs_red} and discuss our findings with respect to {\it Spitzer's}
photometric stability in \S \ref{repeatability}.  The analysis of GU Boo's
light curve is described in \S \ref{analysis}.  We probe for the existence of
an infrared excess in GU Boo's spectral energy distribution in \S
\ref{flux_comparison}.  In \S \ref{other_lightcurves}, we show light curves of
other well sampled objects in the field along with a brief summary of their
respective properties, and we summarize and conclude in \S \ref{conclusions}.


\section{{\it Spitzer} Observations and Data Reduction}\label{obs_red}


\subsection{Observations}\label{observations}

We used the MIPS 24\micron\ array aboard the {\it Spitzer Space Telescope}
\citep{wrl04} to observe GU Boo in February and April of 2006, as
outlined in Table \ref{table1_AORs}. The MIPS 24\micron\ array (MIPS-24), is a
Si:As detector with 128 $\times$ 128 pixels, an image scale of 2.55"
pixel$^{-1}$, and a field of view of 5.4' $\times$ 5.4' \citep{rye04}.  Our
exposures were obtained using the standard MIPS 24$\mu$m small field
photometry pattern, which consists of four cardinal dither positions located
approximately in a square with 2 arcmin on a side.  For two of these cardinal
positions, there are four smaller subposition dithers (offset by $\sim$ 10
arcsec), and for the other two cardinal positions, there are three such
subpositions. This results in a dither pattern in which {\it Spitzer} places
the star at 14 different positions on the array.

Our goal was to observe three independent secondary eclipses of GU Boo: two
consecutive ones and another one several weeks after the first two
\citep{bbc07}.  Of our total of nine of {\it Spitzer's} Astronomical
Observation Requests (AORs), three were used for each secondary eclipse event
(see Table \ref{table1_AORs}).  Each AOR contained eight
exposures\footnote{The term ``exposure'' here can be thought of as a cycle of
observations, but since the word ``cycle'' is reserved for another unit of
{\it Spitzer} data collection, it carries the somewhat misleading name
``exposure''.} with 36 individual Basic Calibrated Data (BCD) frames each. The
first BCD in each exposure is 9s long, the subsequent 35 are 10s long. The
first two BCDs of every exposure were discarded due to a ``first frames
effect''. This procedure left 34 BCDs per exposure, 272 BCDs per AOR, 816 BCDs
per secondary eclipse event, and 2448 BCDs for the entire project (all 10s
exposure time).

For background information on {\it Spitzer} and MIPS, we refer the reader to
the Spitzer Observer's Manual, obtainable at
\url{http://ssc.spitzer.caltech.edu/documents/som/}. For information
specifically related to MIPS data reduction, please consult the MIPS Data
Handbook (MDH -- \url{http://ssc.spitzer.caltech.edu/mips/dh/}) and
\citet{gre05}.

\begin{deluxetable}{ccccc}
\tablecolumns{5}
\tabletypesize{\scriptsize}
\tablewidth{0pc}
\tablecaption{Spitzer MIPS-24 observations of GU Bo\"otis \label{table1_AORs}}
\tablehead{
    \colhead{Date (2006)} &         \colhead{MIPS Campaign} &
    \colhead{Obs. Set} &
    \colhead{AORs} &          \colhead{Exposures\tablenotemark{a}}
} 
\startdata
Feb 20 &   {\it MIPS006500} & 1 & {\it 16105472} &        860 \\
           &     {\it } & & {\it 16105216} &            \\
           &     {\it } & & {\it 16104960} &            \\
Feb 21 &   {\it MIPS006500} & 2 & {\it 16104704} &        860 \\
           &     {\it } & & {\it 16104448} &            \\
           &     {\it } & & {\it 16104192} &            \\
Apr 01 &   {\it MIPS006700} & 3 & {\it 16103936} &        860 \\
           &     {\it } & & {\it 16103680} &            \\
           &     {\it } & & {\it 16103424} &            \\
\enddata
\tablecomments {Two consecutive secondary eclipses were observed in observing
sets 1 and 2, and a third secondary eclipse event six weeks later in observing
set 3.}
\tablenotetext{a}{10 seconds per exposure.}
\end{deluxetable}




\subsection{Data Processing and \textsf{mopex} $/$ \textsf{apex} 
Reduction} \label{reductions}


\subsubsection{Mosaicing}
\label{mosaicing}

The MIPS-24 data are provided by the {\it Spitzer Archive} in the
(flatfielded) BCD format. We applied further post-processing to these data in
order to correct for small scale artifacts, in particular using
IRAF's\footnote{IRAF is distributed by the National Optical Astronomy
Observatory, which is operated by the Association of Universities for Research
in Astronomy, Inc, under cooperative agreement with the National Science
Foundation.} CCDRED package to remove the weak ``jailbar'' features in the
images (as described in the MDH).

\begin{figure}
\plotone{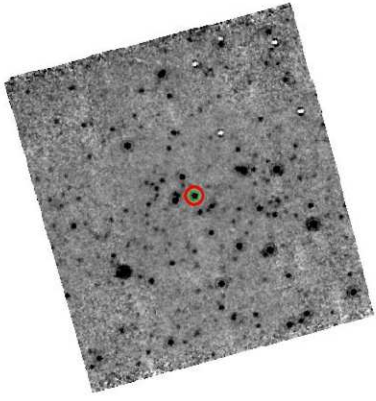} 
\caption{\label{mosaic_AOR} A {\it Spitzer} MIPS 24$\mu$m mosaic of 
GU Boo (marked with a circle at the center of the image). This mosaic was
created using all 272 frames in one AOR and is about 8 arcmin on a side. North
is up, east is to the left. The change in noise structure as a function of
position is due to different effective exposure times (only the inner $\sim$
3' $\times$ 3' were covered by all 272 BCD frames). The white specks in the NW
corner are flatfielding residuals, caused by a fleck of paint or dust grain on
the pickoff mirror (from {\it Spitzer's} launch), imaged at the four cardinal
dither positions (see \S \ref{observations}).}
\end{figure}

The {\it Spitzer} software package \textsf{mopex} \citep{mk05,mm05} was used
to co-add the individual MIPS BCD frames into mosaics of 17 frames, using overlap
correction and outlier rejection in the process. The choice of 17 frames was
made to balance three aspects:

\begin{enumerate}
\item We want to obtain sufficient signal-to-noise ratio (SNR) for measured 
flux densities in the combined images and subsequent data points in the light
curves (SNR $>$ 10 for GU Boo; see Table \ref{tab3_obj_stats}).
\item We need to maintain a sufficiently high effective observing
cadence to temporally resolve elements of GU Boo's light curve for fitting
purposes.
\item We do not want to be forced to combine frames from different exposures
into a single light curve data point (see \S \ref{observations}). 
\end{enumerate}
The interpolated, remapped mosaics have a pixel scale of 2.45''
pixel$^{-1}$. We show in Figure \ref{mosaic_AOR} the MIPS-24 field of view of
GU Boo.


\subsubsection{Photometry}

For photometric reductions of the mosaiced images, we utilized the
\textsf{apex} component of \textsf{mopex} to perform point-source extraction
as described in \citet{mm05}\footnote{Also see information on \textsf{apex} at
\url{http://ssc.spitzer.caltech.edu/postbcd/apex.html} and the User's Guide
at \url{http://ssc.spitzer.caltech.edu/postbcd/doc/apex.pdf}}. This step
included background subtraction of the images, and the fitting of a resampled
point response function (PRF). In order to match the PRF centroid as closely
as possible to the centroid of the stellar profile, the first Airy ring is
initially subtracted from the stellar profiles, and the source detection
happens on the resulting image. Photometry of the detected sources is then
performed on the original images.

For single frame photometry on mosaiced images, \textsf{apex} provides the
option of using a template model PRF produced by the analysis of $\sim$ 20
bright, isolated stars in the {\it Spitzer Archive}, or using one's own data
to create a PRF. It furthermore allows for a resampling factor in both x and y
directions. The scatter in our light curves was minimized when using the model
PRF provided by the {\it Spitzer Science Center}, oversampled by a factor of 4
in both x and y directions. Using the PRF created from our own data or
sampling any PRF to a higher resolution resulted in noticeably larger root
mean square (rms) dispersion in our light curves, most likely due to
systematic errors introduced in the low SNR regime of our data (see Table
\ref{tab3_obj_stats}).

We note that, currently, \textsf{apex} only provides the option of using a
synthetic PRF \citep[Tiny
Tim\footnote{\url{http://ssc.spitzer.caltech.edu/archanaly/contributed/stinytim.}};][]{krist93}
for photometry on individual BCD frames (i.e., not mosaiced), as applied by
\citet{dsr05} and \citet{rhs06}.  Since the flux density of our target star
($\sim$ 600$\mu$Jy) is so much lower than HD 209458 \citep[$\sim$22mJy;][]{dsr05}, our SNR regime did not allow for performing photometry on
single BCD frames.


\subsubsection{Background Fluctuations}
\label{background_fluctuations}

The \textsf{apex} error analysis is described in the \textsf{apex} User's
Guide, and parts of it can be found in \citet{mml02,ml05,mm05}. We briefly
summarize the general idea here. Errors in the photometry are dominated by the
statistical background fluctuations in the images. These fluctuations are
calculated per pixel by estimating the Gaussian noise inside a sliding window
whose size is defined by the user (45 $\times$ 45 interpolated pixels in this
case). Thus, \textsf{apex} produces ``noise tiles'' for the computation of the
SNR of the point sources in the corresponding mosaiced image tiles (see column
7 in Table \ref{tab3_obj_stats}).

\begin{figure}
\includegraphics[angle=-90, width=\linewidth]{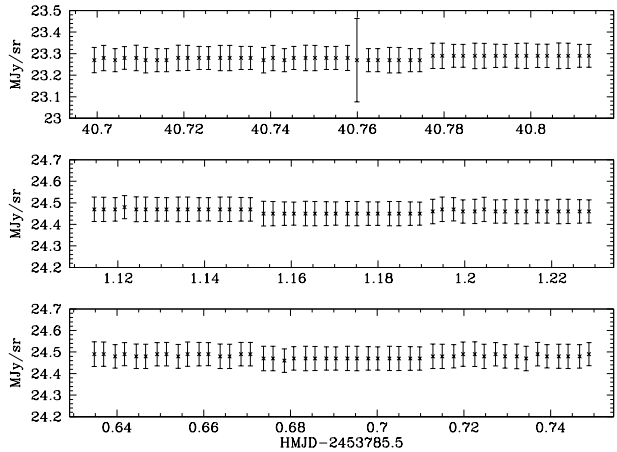}  
\caption{\label{background} Temporal fluctuations of the image background 
measured in surface brightness units of MJy/sr for the three observing sets
(\ref{table1_AORs}). The lower, middle, and upper panel correspond to
observing set 1, 2, and 3, respectively. Note that the background level is
similar between observing sets 1 and 2 ($24.48\pm0.009$ MJy/sr and
$24.46\pm0.009$ MJy/sr, respectively), but lower for observing set 3
($23.28\pm0.008$ MJy/sr). The typical fluctuations within a given image
(represented by the error bars) are around 0.05--0.06 MJy/sr. The large error
bar for the one data point in observing set 3 is caused by a cosmic ray.}
\end{figure}

To provide an estimate of the background fluctuations from image to image, we
show in Fig. \ref{background} the surface brightness for every image in the
three observing sets. These estimates were obtained by calculating the median
surface brightness level for the inner 90\% of the image (in area). The error
bars correspond to the standard deviation about this median over the same
area. Note that the fluctuations of the background within observing sets are
very small, but they are different for the temporally offset observing set 3
(see Table \ref{table1_AORs}). Surface brightness values are given in the {\it
Spitzer} native units of MJy per steradian. The surface brightness values in
the cores of the brightest objects in the field are typically 1--1.5 MJy/sr
above the background level (23--25 MJy/sr). A linear fit (weighted by the
standard deviation values of the data points) to observing sets 1 \& 2 returns
a slope of $-0.033 \pm 0.023$ MJy/sr/day. The same fit for all three observing
sets produced a statistically consistent slope of $-0.030 \pm 0.001$
MJy/sr/day, indicating a smoothly decreasing background level over the course
of our observations (Table \ref{table1_AORs}).

The {\it Spitzer} tool {\textsf
Spot}\footnote{\url{http://ssc.spitzer.caltech.edu/propkit/spot/index.html}}
predicts the surface brightness of the Zodiacal background in {\it Spitzer}
images. \citet{hhl06}, for instance, used this model to correct for ostensible
variations in instrument sensitivity over a period of around four days. We
compared our background estimates to the predictions in {\textsf Spot} and
found that the model underestimates our measurements by 3.5--4.0 MJy/sr. One
potential reason for any offset between observed and estimated backgrounds is
that {\textsf Spot} calculates a monochromatic background, whereas the
measured background is integrated over the wavelength passband and sensitivity
function of the MIPS-24 detector.  Using {\textsf Spot}, we calculated
background estimates in 6 hour increments from 2006 Feb 20 00:00:00 to 2006
Feb 22 00:00:00, and from 2006 Apr 01 00:00:00 to 2006 Apr 02 00:00:00. The
slopes of the Feb backgrounds and the Feb--Apr backgrounds are -0.022 and
-0.032 MJy/sr/day, respectively; {\textsf Spot} does not provide error
estimates in its predictions.  It thus appears that, within statistical
uncertainties, the behavior of the {\textsf Spot} background model is
consistent with our empirical results (at least for the time scales of our
observations), lending further justification to the approach by \citet{hhl06}.


If, however, the difference in slope between the Spot estimates for just Feb
and Feb--Apr is indeed real but simply not detectable at our temporal
resolution (indicating that the background change with time is not a simple
linear function), any resulting discrepancy between calculated and observed
background values may be attributable to the fact that the model is calculated
for Earth, whereas Spitzer is in an Earth-trailing orbit and thus looking
through different amounts of Zodiacal dust \citep{hhl06}.

\section{Precision and Repeatability of the {\it Spitzer} Photometry}
\label{repeatability}


\subsection{PRF Fitting Versus Aperture Photometry}\label{prf_apphot}

\begin{figure}
\plotone{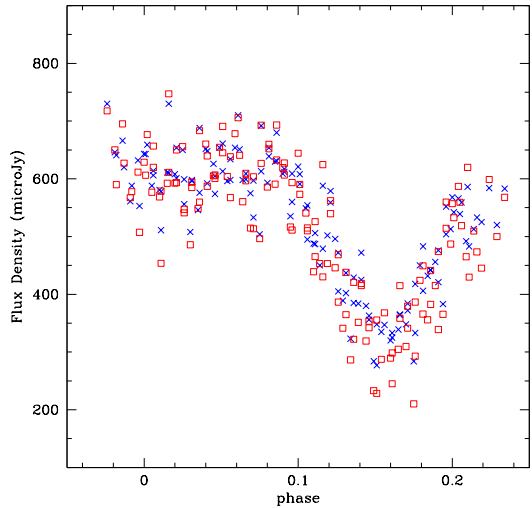} 
\caption{\label{guboo_prf_apphot_comp} Comparison between
PRF photometry (blue crosses) and aperture photometry (red squares) for GU
Boo's phased light curve. Photometric error bars are omitted for the sake of
clarity. The scatter in the flat part of the light curve is smaller for PRF
photometry (49$\mu$Jy) than aperture photometry (59$\mu$Jy), but the median
flux density level of the flat part of the light curve is identical within the
errors for both photometry approaches.}
\end{figure}

To create GU Boo's 24\micron\ light curve, we performed both PRF photometry as
described in \S \ref{reductions} and additionally utilized \textsf{apex's}
option of simultaneously obtaining aperture photometry. Figure
\ref{guboo_prf_apphot_comp} shows the agreement between PRF and aperture
photometry based on an aperture radius of 6'' (with 20--32'' background
annulus) which minimized the rms in the flat part of GU Boo's
phased\footnote{We use the period calculated by \citet{lmr05} (0.488728 days;
see Table \ref{table2_basic_params}) for our phasing throughout this paper.}
light curve. Using information from the MDH, we applied a multiplicative
aperture correction of 1.699 to the photometry.  The median flux density
obtained by PRF photometry for the flat part of the phased light curve is 614
$\pm$ 49$\mu$Jy compared to 608 $\pm$ 59$\mu$Jy for the aperture photometry.
Thus, the absolute median flux density values agree very well for the two
different photometry approaches, but the PRF photometry exhibits smaller
scatter around the median magnitude. We note that the current version of
\textsf{apex} does not calculate photometry errors in the aperture correction,
and the principal reason why we performed aperture correction is to verify the
absolute flux density level of our sources as calculated by PRF fitting. 

\subsection{Absolute Versus Relative Photometry}\label{abs_relphot}

\begin{figure}
\includegraphics[angle=-90, width=\linewidth]{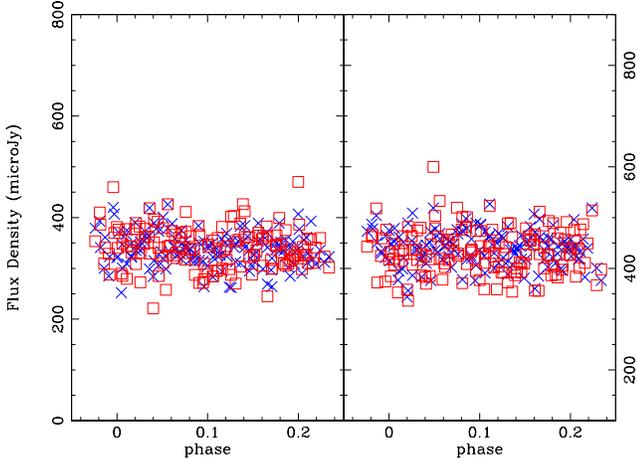}  
\caption{\label{stars_comp} Comparison between PRF photometry (blue crosses) 
and aperture photometry (red squares) for objects 31 (left) and 66 (right).
Photometric error bars are omitted for the sake of clarity. The light curves
are phased to the period of GU Boo for the sake of comparison. The ordinate
scale is the same as Figures \ref{guboo_prf_apphot_comp}, \ref{galaxies_comp},
and \ref{18_lc}--\ref{66_lc}. The flux density values (PRF vs aperture
photometry) are identical within the errors for both objects.}
\end{figure}

\begin{figure}
\includegraphics[angle=-90, width=\linewidth]{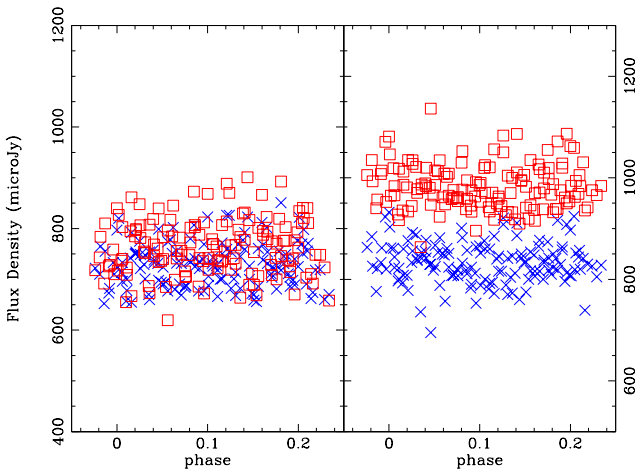}  
\caption{\label{galaxies_comp} Comparison between PRF photometry (blue 
crosses) and aperture photometry (red squares) for objects 18 (left) and 58
(right). Photometric error bars are omitted for the sake of clarity. The light
curves are phased to the period of GU Boo for the sake of comparison. The
ordinate scale is the same as Figures \ref{guboo_prf_apphot_comp},
\ref{stars_comp}, and \ref{18_lc}--\ref{66_lc}. The flux density values (PRF vs
aperture photometry) agree within the errors for object 18, but are discrepant
for object 58.}
\end{figure}

In order to remove statistically correlated noise from GU Boo's light curve,
we performed relative photometry as described in equations 2 and 3 of
\citet{eh01}. We picked comparison objects based on the number of
observational epochs: in order to obtain a relative offset per photometric
data point in GU Boo's light curve (all data points are treated independently
of each other), it is advantageous to use stars with (at least) as many data
points as GU Boo itself. Four objects out of Table \ref{tab3_obj_stats}
fulfill this criterion: numbers 18, 31, 58, and 66 (see Figures
\ref{18_lc}--\ref{66_lc} for their light curves). The cross-referencing in
Table \ref{tab3_obj_stats} shows that objects 31 and 66 are stars, and objects
18 and 58 are galaxies (as are all other objects in the field that we were
able to cross-reference). Note, however, that object 31's closest match in
SDSS \citep{sdss07} and 2MASS \citep{2mass03,2mass06} catalogs is 9'' away
whereas for object 66, the distance to the closest SDSS match was only
0.15''. We originally presumed that, despite the fact that they are galaxies,
objects 18 and 58 would be unresolved at our large pixel size (\S
\ref{observations}) and tested that hypothesis by comparing the flux density
obtained by PRF photometry to that obtained by aperture photometry (6''
aperture; \S \ref{prf_apphot}). Figures \ref{stars_comp} and
\ref{galaxies_comp} shows that our presumption does not hold true for object
58, and we discarded it from our relative photometry procedure.

We find that, by performing differential photometry as outlined above, the
scatter in the flat part of GU Boo's phased light curve reduces by 9.4\% over
the PRF photometry (see Fig. \ref{guboo_prf_apphot_comp}) to 45$\mu$Jy. Light
curve fitting as described in \S \ref{analysis} was performed on the
differential photometry.

\subsection{Intra-Set Versus Inter-Set Photometric Stability}

\begin{figure}
\epsscale{0.7}\plotone{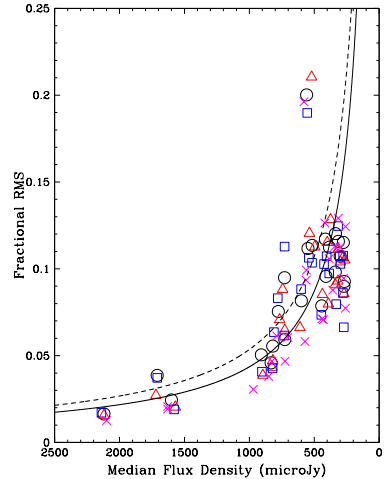} 
\caption{\label{global_rms} 
Median flux density versus fractional rms for the 24 objects that have
photometry for more than 72 out of 144 observational epochs. Triangles,
squares, and crosses represent the data from observings sets 1, 2, and 3,
respectively (see Table \ref{table1_AORs}), to illustrate the repeatability of
{\it Spitzer}/MIPS-24 within individual observing sets. Circles mark the data
points from the combination of all 3 observing sets (to show the inter-set
stability). The data point with the highest fractional rms is GU Boo, due to
its intrinsic variability. When subtracting our fit from its light curve (see
Fig. \ref{lc_fit}), GU Boo's fractional rms falls to 0.081. The solid and
dashed lines indicate MIPS-24's sensitivity for our exposure times as a
function of flux density for low and medium background levels, respectively.}
\end{figure}

\begin{figure}
\includegraphics[angle=-90, width=\linewidth]{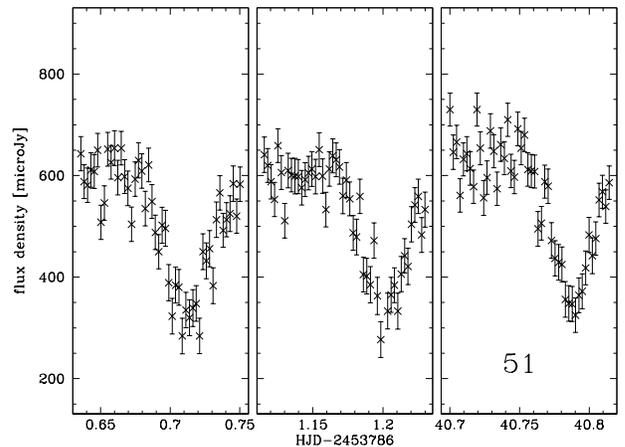}
\caption{\label{lc_guboo} GU Boo's 24$\micron$ light curve, based on absolute
PRF photometry (see \S \ref{prf_apphot} and \S \ref{abs_relphot}).  The three
panels represent the three MIPS-24 observing sets during which the individual
secondary eclipse events were observed. GU Boo is object number 51 in our
numbering system (see Table \ref{tab3_obj_stats}).}
\end{figure}

Figure \ref{global_rms} shows fractional rms values versus median flux
densities for all objects in Table \ref{tab3_obj_stats}. For every object, we
plot fractional rms for each individual observing set as well as for the
three sets combined. Observing sets 1 and 2 were obtained during the
MIPS006500 campaign, observing set 3 during MIPS006700 (Table
\ref{table1_AORs}).  Consistent with the results in \citet{rye04}, we find
that inter-set repeatability of {\it Spitzer's} MIPS-24 is comparable to the
intra-set repeatability, both in terms of median flux density and the rms
scatter of the light curves, despite varying background levels (see \S
\ref{background_fluctuations}).  For objects with a flux density in excess of
1mJy, the rms scatter approaches 1--2 \%, similar to the scatter found for the
brightest sources observed with MIPS-24 in \citet{rye04}. Because of the
intrinsic variability produced by the stellar eclipse, GU Boo has the largest
fractional rms ($\sim$ 0.2). However, when we subtract the fit (see \S
\ref{analysis}) from GU Boo's light curve, the fractional rms falls to 0.081,
consistent with stars of similar median brightness. We show GU Boo's light
curve for the three individual observing sets in Fig. \ref{lc_guboo} and the
phased light curve along with the fit in Fig. \ref{lc_fit}.



In order to compare our rms values to background-limited noise values, we used
{\it Spitzer's} {\textsf
SENS-PET}\footnote{\url{http://ssc.spitzer.caltech.edu/tools/senspet}} to
predict the MIPS-24 sensitivity ($1 \sigma$ above background for 170 seconds
integration time; see \S \ref{mosaicing}) for low and medium background levels
(solid and dashed line in Fig. \ref{global_rms}, respectively).  The
mid-infrared background at the time of observations of GU Boo is 23--24.5
MJy/sr, which is between the typical low and medium background levels used by
{\textsf SENS-PET} (see also \S \ref{background_fluctuations}). We find that
the {\textsf SENS-PET} predictions are consistent with our empirically
determined error estimates.  Except for the variable GU Boo, typical values
for the rms scatter of the light curves (Table \ref{tab3_obj_stats}) are
approximately equal to average photometric measurement uncertainties of
individual data points (see Figs. \ref{18_lc}, \ref{31_lc}, and \ref{66_lc}).

\section{Analysis of GU Boo's Photometric Light Curve}\label{analysis}

We modeled the secondary eclipse observations of GU Boo using the JKTEBOP
code \citep{sms04,szm04}. JKTEBOP is based on the original EBOP code
\citep{pe81,e81}, but with the addition of the Levenberg-Marquardt
optimization algorithm \citep{NR92} to find the best fitting model, and also
the implementation of a Monte Carlo simulation algorithm to determine robust
uncertainties in the fitted parameters \citep{ssm05}.

The orbital period and initial epoch of the primary eclipse were set to the
values given in the ephemeris derived by \citet{lmr05}. We further fixed the
mass ratio and the radius ratio of the stars, as well as the eccentricity of
the system ($e$=0) to the values obtained in that work. We assumed no limb
darkening effects in the light curves, as expected for observations this far
into the infrared \citep[][and references therein]{cdg95,rhs06,cbb07,s07}, and
no significant gravitational darkening or reflection effects, based on the
spherical shape of the stars and the similarity in effective temperatures. All
these are reasonable assumptions, based on the results of the study of GU Boo
at visible wavelengths, and they are, in fact, hard to test in detail, given
the photometric precision of the {\it Spitzer} light curve at this flux
density level.

In the absence of primary eclipse observations, to calculate the luminosity
ratio of the system, we place a further constraint to the fit by fixing the
value of the surface brightness ratio of the stars to J=J2/J1=0.9795. This
value, combined with the adopted radius ratio and the no limb darkening
assumption, gives a luminosity ratio of of $L_{2}/L_{1}$ = 0.9697, which is
consistent with the expected value for GU Boo at 24$\mu$m.

The parameters initially left free in the models were: (1) the fractional sum
of the radii, i. e., $(R_{1} + R_{2})/a$, where $R_{1}$ and $R_{2}$ are the
component radii, and $a$ is the orbital separation, computed from the stellar
masses and the orbital period of the system, (2) the inclination of the orbit
$i$, (3) the amount of third light $L_{3}$, and (4) a phase offset parameter
$\phi$ (to account for small errors in the ephemeris).

\begin{figure}
\plotone{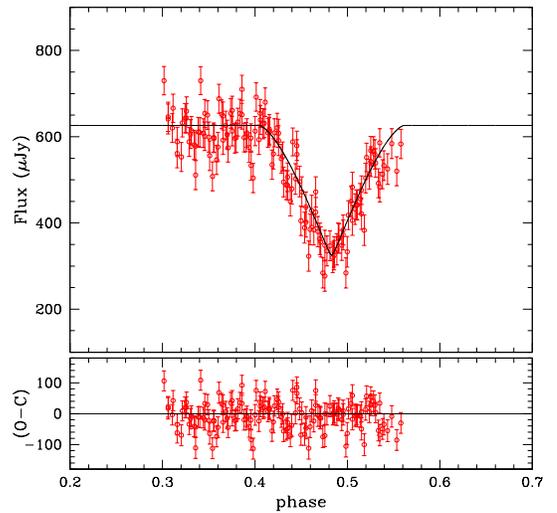}
\caption{\label{lc_fit} Our best fit overlaid on top of the phased 24$\micron$
data of GU Boo derived from our relative photometry of the system. The fit has
a reduced $\chi^2 = 1.7 \pm$ 0.1. The bottom panel shows the residuals around
the fit. The calculated system parameters, which agree well with the results
from the optical study in \citet{lmr05}, are shown in Table
\ref{table2_basic_params}. }
\end{figure}

 
Our best model solution is illustrated in Figure \ref{lc_fit}, with a reduced
$\chi^2$ of 1.7, and a mean fractional error per data point of 9.5\%
(cf. Fig. \ref{global_rms}). Formal errors in the fitted parameters were
derived using the Monte Carlo algorithm implementation in JKTEBOP for a total
of 1000 iterations.  We obtain a radius for the secondary component of $R_{2}$
= 0.66 $\pm$ 0.02 $R_{\sun}$.  Our value of the orbital inclination is $i$ =
89.3 $\pm$ 0.8 degrees.  Both values are slightly larger than the ones
obtained by \citet{lmr05} at optical wavelengths, $R_{2}$ = 0.62 $\pm$ 0.02
$R_{\sun}$ and $i$ = 87.6 $\pm$ 0.2 degrees. The two secondary radius
estimates agree to within random statistical errors (1.4$\sigma$).  In the
case of the inclination, our value is not as well constrained as in the
optical, since we lack a full light curve that includes a primary eclipse.  We
show our estimates for GU Boo's system parameters in Table
\ref{table2_basic_params}.  For the tested third light contribution, we obtain
a value of $L_{3}$ = -0.04 $\pm$ 0.07, consistent with $L_{3}$ = 0.

Finally, we find a phase shift of $\Delta\phi$ = -0.014 $\pm$ 0.001.  This
phase shift is 1.5 times larger than expected from the \citet{lmr05}
($|\Delta\phi|$ = 0.009), but can still be attributed to uncertainties in the
original period estimation. The \citet{lmr05} observations were conducted in
2003, near JD=2452733.  The number of elapsed periods inbetween those
observations and our {\it Spitzer} AORs is about 2150.  The 1-$\sigma$ error
in the \citet{lmr05} period estimate is $2\times 10^{-6}$ days, which
accumulates to 0.0043 days, about 0.009 in phase, in 2150 periods. Thus, the
discrepancy we find corresponds to about 1.5$\sigma$ from the \citet{lmr05}
ephemeris predictions. We estimate that this offset is based on normal
statistiscal errors. An alternative explanation would be that a third body
orbiting the system could cause this shift, but since (1) we show in \S
\ref{flux_comparison} that GU Boo's flux is consistent with its modeled
spectral energy distribution, and (2) we calculate the third light component
to be $L_{3}$ = 0, any such claim would be unsubstantiable with our data.

Equation \ref{ephemeris} shows the updated ephemeris equation of GU Boo by
combining the seven minima in table 5 of \citet{lmr05} with the three new
minima presented in this work. 

\begin{equation}
\label{ephemeris}
T(\mbox{Min I}) = \mbox{HJD}2452723.981327(1)+  0.4887247(8) \cdot E.
\end{equation}

Uncertainty digits are given in parentheses. $E$ represents the number of
elapsed periods since the initial epoch, $T(\mbox{Min I})$ the time of primary
eclipse minimum.

\begin{deluxetable}{lr}
\tablecolumns{2}
\tabletypesize{\scriptsize}
\tablewidth{0pc}
\tablecaption{GU Boo System Parameters\label{table2_basic_params}}
\tablehead{
    \colhead{Parameter} &         \colhead{Value} \\
}
\startdata

Orbital Period (days)\tablenotemark{a} & $0.488728  \pm  0.000002 $ \\
Orbital Eccentricity\tablenotemark{a} &          0 (fixed)\\
Mass Ratio (M2/M1) \tablenotemark{a} & $0.9832  \pm  0.0069 $ \\
Combined out-of-eclipse 24 \micron\ flux ($\mu$Jy) & $614  \pm  49 $ \\
Radius of Secondary Component ($R_{\sun}$) & 0.66 $\pm$ 0.02 (0.62\tablenotemark{a})\\
Orbital Inclination $i$ (degrees) & 89.3 $\pm$ 0.8 (87.6\tablenotemark{a})\\

\enddata
\tablenotetext{a}{\citet{lmr05}}
\end{deluxetable}


\label{basic_params}


\section{Comparison Between Expected and Measured 24\micron\ Flux Density of 
GU Boo}\label{flux_comparison}

In addition to the relative photometry of GU Boo (Fig. \ref{lc_fit}), we also
performed absolute photometry as reported in \S \ref{prf_apphot}. The
24\micron\ flux density of $614 \pm 49\mu$Jy was determined from the median
flux level outside of eclipse.  To test the accuracy of the absolute flux
density level, we show in this section a spectral energy distribution (SED)
model between 0.11 and 35$\micron$, scaled to the optical and near-infrared
(NIR) magnitudes of GU Boo \citep{lmr05}.

\begin{figure}
\includegraphics[angle=90, width=\linewidth]{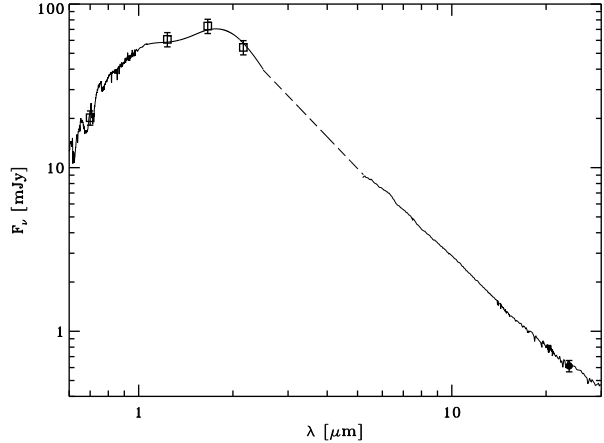}
\caption{\label{guboo_sed} The SED of GU Boo, based on M1V $0.11-2.5$\micron\
optical-NIR templates of \citet{p98} and {\it Spitzer} $5-35$\micron\ IRS
spectra of GL~229A (spectral type M1V; $T_{eff}$ = 3800 K). The dashed line
respresents the interpolation between template and spectra. For details, see
\S \ref{flux_comparison}.}
\end{figure}

The GU Boo system components are two M stars of nearly identical mass,
temperature and radius. For our SED model, we assumed both stellar components
to be M1V stars with effective temperatures of 3800 K
\citep[e.g.,][]{lmr05}. The model SED was constructed from the M1V
$0.11-2.5$\micron\ optical-NIR templates of \citet{p98} and the {\it Spitzer}
$5-35$\micron\ Infrared Spectrograph \citep[IRS; ][]{hrc04} spectra of
GL~229A, an M1V (3800 K) star \citep{crm06}. To build the SED model
(Fig. \ref{guboo_sed}), the M1V optical-NIR template was scaled to GU Boo's
optical-NIR flux densities based on table 1 in \citet{lmr05}. To connect the
{\it Spitzer} IRS spectrum to the optical-NIR template, we fit a power law of
the form $F_\nu \propto \nu^{n}$ (see dashed line in Fig. \ref{guboo_sed}) to
the IRS spectrum. We found the best-fit exponent to the power law to be
$n=1.9$.  The IRS spectrum and the power law (extrapolated to 2.4$\micron$)
were then scaled to the red edge of the optical-NIR template. The slope of the
power law was maintained to ensure a continuous transition between the
optical-NIR template and the IRS spectrum (see Fig. \ref{guboo_sed}).  Note
that only the optical and NIR flux densities were used to scale the SED model;
i.e., the scaling does not utilize the 24\micron\ data point.

The SED model predicts a mid-infrared flux density for GU Boo of
$F_{\nu}(24\mu{\rm m}) \approx 650\mu$Jy.  The measured 24\micron\ flux
density of GU Boo ($614 \pm 49\mu$Jy) is within 1$\sigma$ of the predicted
flux density, agreeing remarkably well with the simple SED model presented
here. We conclude that the stellar components are solely responsible for the
mid-infrared emission of GU Boo.

The few M- and K-dwarf DEB systems studied to date (GU Boo included) reveal
that many of the binary components have larger radii (by 10-20\%) and cooler
effective temperatures (by 100 K to hundreds of K) than predicted by stellar
evolutionary models \citep[e.g.,][]{tr02,ribas03,lmr05,lm07}. Magnetic
activity and metallicity can account for the radius discrepancy
\citep{lm07} and, in principle, also for the temperature discrepancy.  An
alternative explanation for the temperature discrepancy, however, is the
presence of dusty material around the systems.  The excellent agreement of our
observed mid-infrared flux density with the model SED suggests that there is
little, if any, (warm) circumstellar dust in GU Boo, likely ruling out
circumstellar dust as a viable explanation for discrepancies with the stellar
evolutionary models.


\section{Light Curves of Selected Objects in the Field of GU Boo}
\label{other_lightcurves}

\begin{figure}
\includegraphics[angle=-90, width=\linewidth]{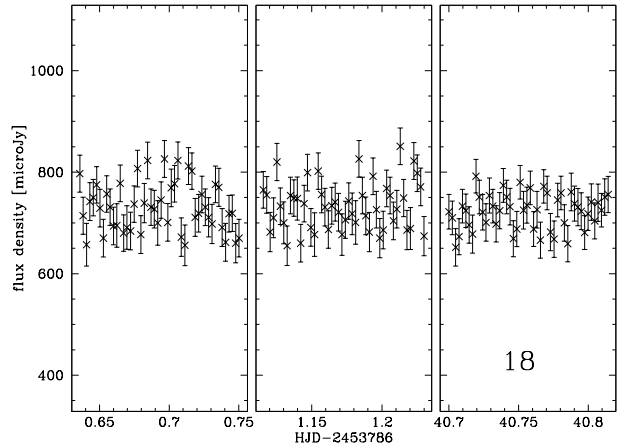}
\caption{\label{18_lc} The MIPS-24 light curves of object 18 (a galaxy) in
the field of GU Boo. The three panels represent the three MIPS-24 observing
sets. For parameters, see Table \ref{tab3_obj_stats}.}
\end{figure}

\begin{figure}
\includegraphics[angle=-90, width=\linewidth]{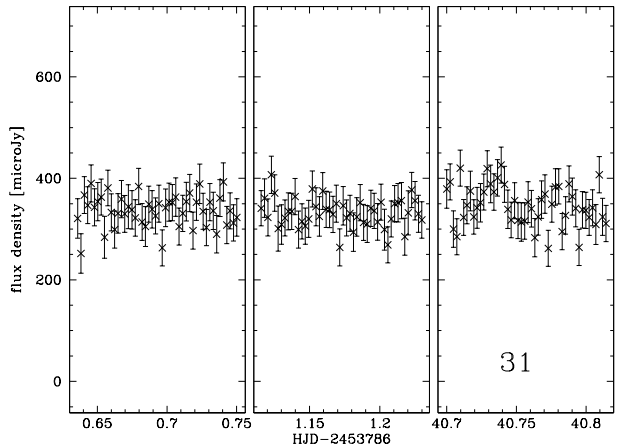}
\caption{\label{31_lc}The MIPS-24 light curves of object 31 (possibly a star) in
the field of GU Boo. The three panels represent the three MIPS-24 observing
sets. For parameters, see Table \ref{tab3_obj_stats}. If the cross
referencing in \S \ref{abs_relphot} is correct, object 31 is an M3III giant
(see text in \S \ref{other_lightcurves}).}
\end{figure}

\begin{figure}
\includegraphics[angle=-90, width=\linewidth]{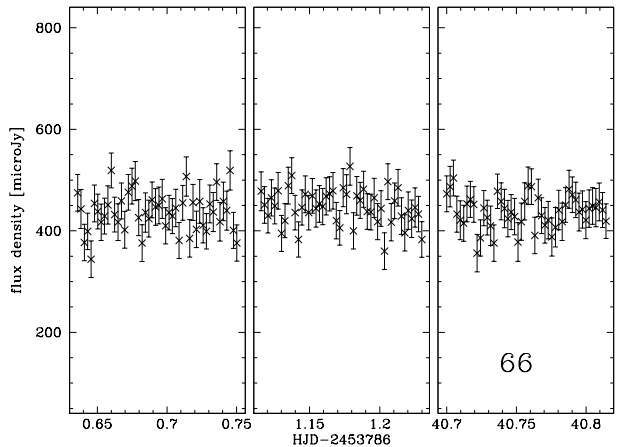}
\caption{\label{66_lc}The MIPS-24 light curves of object 66 (a star) in
the field of GU Boo. The three panels represent the three MIPS-24 observing
sets. For parameters, see Table \ref{tab3_obj_stats}. Our SED fitting
indicates this to be an A2V dwarf (see text in \S \ref{other_lightcurves}).}
\end{figure}

In this Section, we present a brief summary of selected other light curves in
the field of GU Boo, along with basic determination of spectral types of the
objects identified as stars (see \S \ref{abs_relphot}). We limit our selection
to the three objects that were used to perform the relative photometry (see \S
\ref{abs_relphot}).  Figures \ref{18_lc}--\ref{66_lc} display these light
curves. They are all on the same scale with different zeropoints. Parameters
for all objects with at least 72 out of the 144 epochs are listed in Table
\ref{tab3_obj_stats}. We do not show light curves for the rest of the field
objects since they can essentially be described as flat lines with some
scatter around the median magnitude, which is characterized by the values in
Table \ref{tab3_obj_stats}.

Spectral typing for the two stars (objects 31 and 66) was attempted by means
of SED fitting of photometry available in the literature: both objects have
Sloan DSS $ugriz$ \citep{sdss07} data points, and star 31 additionally has
2MASS $JHK_s$ \citep{2mass03,csd03} and Johnson $RI$ \citep{mlc03} magnitudes
available for it.  SED fits were performed using the {\tt sedFit} program
discussed in \S 3.1 of \citet{bcb07}. The best SED match for star 31 is an
M3III giant ($\chi^2_{reduced} \sim 1.6$), whereas star 66's SED was found to
be consistent with an A2V dwarf ($\chi^2_{reduced} \sim 0.9$).  Note that one
assumption we make here is that the cross referencing for object 31 is
correct, despite the large distance from its closest matches in the SDSS and
2MASS catalogs (Table \ref{tab3_obj_stats}).

\section{Summary and Conclusions}\label{conclusions}

We used MIPS-24 onboard the {\it Spitzer Space Telescope} to obtain
time-series photometry of the M-dwarf DEB GU Boo. Our observations cover three
secondary eclipse events, two consecutive ones and an additional event six
weeks later. Analysis of the photometry shows that the flux density values for
aperture photometry and PRF photometry agree, and that the PRF photometry
produces smaller scatter in the light curve. This scatter can be further
reduced by performing relative photometry based on three comparison objects in
the field. We find that the repeatability of MIPS-24 photometry is consistent
over all temporal scales we sampled: within an observing set and on time
scales of 24 hours and six weeks.

Our mid-IR analysis of GU Boo's light curve is less affected by stellar
surface features than its optical counterpart. The results we produce show
very good agreement with the previously obtained system parameters based on
optical and near-IR work. A comparison between GU Boo's flux density and its
model SED based on stellar templates and IRS spectra shows no IR excess,
leading us to the conclusion that no warm circumstellar dust is present in the
system. 

Finally, light curves of other objects in the field indicate that the
photometric stability of {\it Spitzer's} MIPS-24 is comparable over short
(hours to days) and long (weeks) time scales, despite fluctuations in the
image mid-IR background on time scales of weeks.


\acknowledgements

We gratefully acknowledge the allocation of {\it Spitzer} Director's
Discretionary Time (DDT) for this project. We furthermore thank D. Frayer,
S. Carey, J. Colbert, and P. Lowrance for providing valuable insight into the
mysterious world of \textsf{mopex} and \textsf{apex}, as well as the anonymous
referee for the thorough study of the manuscript and some very insightful
comments and suggestions that significantly improved the quality of this
publication. Thanks also to J. Southworth for useful clarifications on the use
of JKTEBOP, and to M. Cushing for providing the IRS spectrum of GL 229A in
electronic format.  This research has made use of NASA's Astrophysics Data
System. This publication makes use of data products from the Two Micron All
Sky Survey, which is a joint project of the University of Massachusetts and
the Infrared Processing and Analysis Center/California Institute of
Technology, funded by NASA and the NSF.  Funding for the SDSS and SDSS-II has
been provided by the Alfred P. Sloan Foundation, the Participating
Institutions, the National Science Foundation, the U.S. Department of Energy,
NASA, the Japanese Monbukagakusho, the Max Planck Society, and the Higher
Education Funding Council for England. The SDSS Web Site is
\url{http://www.sdss.org/}.  The SDSS is managed by the Astrophysical Research
Consortium for the Participating Institutions. The Participating Institutions
are the American Museum of Natural History, Astrophysical Institute Potsdam,
University of Basel, University of Cambridge, Case Western Reserve University,
University of Chicago, Drexel University, Fermilab, the Institute for Advanced
Study, the Japan Participation Group, Johns Hopkins University, the Joint
Institute for Nuclear Astrophysics, the Kavli Institute for Particle
Astrophysics and Cosmology, the Korean Scientist Group, the Chinese Academy of
Sciences (LAMOST), Los Alamos National Laboratory, the Max-Planck-Institute
for Astronomy (MPIA), the Max-Planck-Institute for Astrophysics (MPA), New
Mexico State University, Ohio State University, University of Pittsburgh,
University of Portsmouth, Princeton University, the United States Naval
Observatory, and the University of Washington.


\bibliography{spitzer}


\clearpage

\begin{landscape}

\begin{deluxetable}{rrrcccrrrrrr}
\tabletypesize{\scriptsize}
\setlength{\tabcolsep}{0.02in}
\tablecolumns{11}
\tablecaption{Basic Parameters of Objects in the Field of GU Boo \label{tab3_obj_stats}}
\tablewidth{0pt}
\tablehead{
\colhead{ID} &         \colhead{$\alpha_{2000}$} &        \colhead{$\delta_{2000}$} &
\colhead{2MASS ID} &    \colhead{SDSS ID} &  \colhead{SDSS type} &
\colhead{SNR} & \colhead{flux ($\mu$Jy)} & 
\colhead{flux$_{1}$ ($\mu$Jy)} & 
\colhead{flux$_{2}$ ($\mu$Jy)} & 
\colhead{flux$_{3}$ ($\mu$Jy)} & 
\colhead{epochs}} 
\startdata
         2 & 15:21:43.27 & 33:52:53.00 & -- & J152143.31+335252.6 & Galaxy &
         11.2 & 729 $\pm$ 69.19 & 743 $\pm$ 65.53 & 729 $\pm$ 82.25 & 704
         $\pm$ 42.61 & 114 \\

         3 & 15:21:41.20 & 33:53:10.99 & -- & J152141.26+335310.1 & Galaxy &
         9.0 & 549 $\pm$ 61.34 & 537 $\pm$ 64.70 & 542 $\pm$ 57.65 & 560 $\pm$
         55.56 & 72 \\

         7 & 15:21:57.76 & 33:52:27.17 & -- & J152157.77+335226.9 & Galaxy &
         6.6 & 414 $\pm$ 48.48 & 400 $\pm$ 46.08 & 424 $\pm$ 43.55 & 418 $\pm$
         52.85 & 129 \\

        12 & 15:21:51.50 & 33:53:55.18 & -- & J152151.61+335355.1 & Galaxy &
        13.9 & 820 $\pm$ 45.50 & 819 $\pm$ 38.65 & 811 $\pm$ 51.52 & 827 $\pm$
        38.69 & 127 \\

        16 & 15:21:45.61 & 33:54:46.99 & 15214563+3354466 &
        J152145.64+335446.4 & Galaxy & 31.4 & 1710 $\pm$ 66.06 & 1720 $\pm$
        46.81 & 1710 $\pm$ 63.68 & 1630 $\pm$ 31.41 & 120 \\

        17 & 15:22:00.15 & 33:54:00.34 & -- & -- & -- & 7.0 & 383 $\pm$ 43.07
        & 374 $\pm$ 48.01 & 384 $\pm$ 37.40 & 384 $\pm$ 40.65 & 134 \\

        18\tablenotemark{a} & 15:21:49.41 & 33:54:53.36 & -- &
        J152149.41+335452.4 & Galaxy & 13.8 & 727 $\pm$ 43.08 & 727 $\pm$
        47.02 & 731 $\pm$ 45.03 & 725 $\pm$ 33.90 & 144 \\

        19 & 15:21:40.34 & 33:55:27.20 & 15214031+3355259 &
        J152140.34+335526.0 & Galaxy & 39.9 & 2120 $\pm$ 34.89 & 2120 $\pm$
        33.17 & 2140 $\pm$ 36.68 & 2100 $\pm$ 26.63 & 72 \\

        29 & 15:21:40.75 & 33:55:54.64 & -- & -- & -- & 7.9 & 410 $\pm$ 39.22
        & 391 $\pm$ 31.21 & 405 $\pm$ 43.46 & 430 $\pm$ 30.42 & 72 \\

        31\tablenotemark{a} & 15:21:52.83 & 33:55:13.38 &
        15215249+3355046\tablenotemark{c} &
        J152152.53+335504.9\tablenotemark{c} & Star? & 6.6 & 338 $\pm$ 33.17 &
        338 $\pm$ 30.70 & 331 $\pm$ 26.36 & 342 $\pm$ 38.59 & 144 \\

        37 & 15:21:46.94 & 33:55:49.89 & -- & -- & -- & 12.1 & 599 $\pm$ 48.87
        & 610 $\pm$ 40.35 & 601 $\pm$ 53.10 & 572 $\pm$ 33.25 & 116 \\

        44 & 15:21:54.15 & 33:55:44.33 & -- & J152154.21+335544.1 & Galaxy &
        6.8 & 336 $\pm$ 40.41 & 313 $\pm$ 34.81 & 316 $\pm$ 39.37 & 357 $\pm$
        31.34 & 136 \\

        46 & 15:21:52.71 & 33:55:53.94 & -- & -- & -- & 6.2 & 300 $\pm$ 32.19
        & 298 $\pm$ 30.99 & 299 $\pm$ 30.75 & 300 $\pm$ 33.70 & 129 \\

        51\tablenotemark{b} & 15:21:54.80 & 33:56:09.35 & 15215482+3356088 &
        J152154.83+335608.9 & Star & 11.0 & 559 $\pm$ 111.90 & 520 $\pm$
        109.45 & 554 $\pm$ 105.12 & 579 $\pm$ 113.62 & 143 \\

        55 & 15:21:44.87 & 33:56:48.86 & -- & J152144.89+335648.4 & Galaxy &
        31.8 & 1600 $\pm$ 39.13 & 1570 $\pm$ 32.05 & 1580 $\pm$ 30.08 & 1630
        $\pm$ 33.78 & 72 \\

        58 & 15:21:48.90 & 33:56:48.06 & 15214889+3356478 &
        J152148.87+335647.3 & Galaxy & 16.7 & 832 $\pm$ 38.17 & 826 $\pm$
        35.79 & 821 $\pm$ 35.08 & 851 $\pm$ 32.40 & 144 \\

        65 & 15:21:59.48 & 33:56:20.17 & -- & -- & -- & 5.8 & 276 $\pm$ 31.80
        & 280 $\pm$ 29.77 & 277 $\pm$ 29.81& 259 $\pm$ 32.21 & 115 \\

        66\tablenotemark{a} & 15:21:56.32 & 33:56:37.66 & -- &
        J152156.31+335637.6\tablenotemark{d} & Star & 9.0 & 442 $\pm$ 34.76 &
        437 $\pm$ 37.28 & 446 $\pm$ 32.79 & 441 $\pm$ 31.39 & 144 \\

        70 & 15:21:58.89 & 33:56:33.23 & -- & -- & -- & 5.6 & 268 $\pm$ 24.85
        & 271 $\pm$ 23.22 & 279 $\pm$ 24.02 & 260 $\pm$ 20.10 & 73 \\

        74 & 15:21:49.62 & 33:57:20.07 & -- & J152149.59+335718.8 & Galaxy &
        5.2 & 272 $\pm$ 24.46 & 265 $\pm$ 27.86 & 272 $\pm$ 18.05 & 273 $\pm$
        25.82 & 77 \\

        84 & 15:22:03.07 & 33:57:24.75 & 15220305+3357239 &
        J152203.05+335724.2 & Galaxy & 17.2 & 908 $\pm$ 45.81 & 896 $\pm$
        34.68 & 903 $\pm$ 36.81 & 969 $\pm$ 29.64 & 120 \\

        85 & 15:21:57.43 & 33:57:45.84 & -- & -- & -- & 5.9 & 314 $\pm$ 36.37
        & 318 $\pm$ 29.56 & 303 $\pm$ 32.52 & 315 $\pm$ 40.68 & 105 \\

        90 & 15:21:58.86 & 33:59:14.34 & -- & J152158.83+335915.2 & Galaxy &
        11.8 & 777 $\pm$ 58.61 & 766 $\pm$ 54.21 & 781 $\pm$ 64.90 & 789 $\pm$
        48.39 & 120 \\

        94 & 15:21:51.67 & 33:59:58.95 & -- & -- & -- & 7.3 & 516 $\pm$ 58.57
        & 496 $\pm$ 55.89 & 517 $\pm$ 53.49 & 563 $\pm$ 52.59 & 93 \\

\enddata

\tablecomments{Parameters for all objects in the field of view of GU Boo with
at least 72 epochs (half the total number). Listed are source ID, position,
2MASS \citep{2mass03,2mass06} and SDSS \citep{sdss07} cross-referenced IDs,
and SDSS type (if available). Flux densities are median values, rms denotes
the scatter around the median.  SNR (column 7) represents the average SNR per
data point in the light curve (see \S \ref{reductions}).  Flux density in
column 7 indicates the overall median value, whereas the flux densities
flux$_{1}$, flux$_{2}$, and flux$_{3}$, (columns 9 -- 11) show the values for
the 1st, 2nd, and 3rd observing sets (Table \ref{table1_AORs}),
respectively. The last column is the total number of data points for the
object.}

\tablenotetext{a}{Used as comparison object for differential photometry (\S
\ref{abs_relphot}).} 
\tablenotetext{b}{GU Boo.}
\tablenotetext{c}{This is the closest match in both SDSS and 2MASS catalogs,
with a distance of $\sim$ 9'', rendering the cross referencing somewhat
uncertain.}
\tablenotetext{d}{The distance between object 66 and the best SDSS match is
around 0.15''.}

\end{deluxetable}

\clearpage

\end{landscape}


\end{document}